\documentclass[]{elsarticle}
\usepackage[utf8]{inputenc}
\usepackage{amsmath,mathtools}
\usepackage{tensor}
\usepackage{amssymb}
\usepackage{graphicx}
\usepackage{lipsum}  
\usepackage{verbatim}

\usepackage{braket}
\usepackage{setspace}
\usepackage{comment}

\newcommand{\E}{\tilde{E}}
\newcommand{\f}{\phi}

\begin{document}
\title{Generalized Ashtekar variables for  Palatini $f(\mathcal{R})$ models}

 \author[1]{Flavio Bombacigno}
 \ead{flavio.bombacigno@ext.uv.es}
 
 \author[2,3,4]{Simon Boudet\corref{cor1}}
 \ead{simon.boudet@unitn.it}

 \author[4,5]{Giovanni Montani}
 \ead{giovanni.montani@enea.it}
 
\address[1]{Departament
de Física Teòrica and IFIC, Centro Mixto Universitat de València - CSIC. Universitat de València, Burjassot-46100, València,
Spain}

\address[2]{Department of Physics, University of Trento, Via Sommarive 14, 38123 (Trento), Italy}

\address[3]{TIFPA - INFN, Via Sommarive 14, 38123 Povo (TN), Italy}

\address[4]{Physics Department, ``Sapienza'' University of Rome, P.le Aldo Moro 5, 00185 (Roma), Italy}

\address[5]{ENEA, Fusion and Nuclear Safety Department, C. R. Frascati,
 	Via E. Fermi 45, 00044 Frascati (Roma), Italy}

\cortext[cor1]{Corresponding author}

\begin{abstract}
We consider special classes of Palatini $f(\mathcal{R})$ theories, featured by additional Loop Quantum Gravity inspired terms, with the aim of identifying a set of modified Ashtekar canonical variables, which still preserve the $SU(2)$ gauge structure of the standard theory. In particular, we allow for affine connection to be endowed with torsion, which turns out to depend on the additional scalar degree affecting Palatini $f(\mathcal{R})$ gravity, and in this respect we successfully construct a novel Gauss constraint. We analyze the role of the additional scalar field, outlining as it acquires a dynamical character by virtue of a non vanishing Immirzi parameter, and we describe some possible effects on the area operator stemming from such a revised theoretical framework. Finally, we compare our results with earlier studies in literature, discussing differences between metric and Palatini approaches. It is worth noting how the Hamiltonian turns out to be different in the two cases. The results can be reconciled when the analysis is performed in the Einstein frame.
\end{abstract}


\maketitle

\section{Introduction}
The most successful approach to the problem of quantizing gravity is, up to now, the so called Loop Quantum Gravity (LQG) theory \cite{Cianfrani2014,Rovelli2004,Thiemann2007}. This formulation, of course, still contains a number of unsolved issues, like the implementation of the quantum dynamics via the scalar constraint, the construction of a classical limit and the ambiguity in the meaning and value of the so-called Immirzi parameter \cite{Rovelli1998,Gambini1999,
Chou2005,Perez2006,Mercuri2006,Date2009,
Mercuri2009,Nicolai2005}. Nonetheless, the great interest for LQG is due to the possibility of constructing a kinematic Hilbert space for the quantum theory, resulting in geometrical operators like areas and volumes, endowed with discrete spectra \cite{Rovelli1995,Ashtekar1997,Lewandowski1997}. By other words, LQG is able to introduce space discretization starting from a classical Lagrangian for the gravitational field \cite{Cianfrani2014}, with the quantum theory just relying on the pre-metric concept of graph. That is achieved via  Ashtekar-Barbero-Immirzi variables \cite{Ashtekar1986,Ashtekar1987, BarberoG.1995,Immirzi1997}, which allow an Hamiltonian formulation of gravity by close analogy with non-Abelian gauge theories: the constraint associated to local spatial rotations can be put in the form of a standard Gauss constraint for the $SU(2)$ group. When tetrads and spin connections are considered independent fields, however, it is necessary to add to the Palatini action new terms, the Holst or Nieh-Yan contributions, which do not affect the classical dynamics (the former is vanishing on half-shell, where the equation for the connection holds, and the latter a pure topological term \cite{Nieh1982,Holst1996,Soo1999,Nieh2007,
Mercuri2008,Banerjee2010}). In both cases, then, we deal with a restatement of General Relativity, suitable for loop quantization, which admits Einstein equations as classical limit. 
\\ \indent In this respect, the recent interest for $f(\mathcal{R})$ modifications of General Relativity \cite{Sotiriou2010, Nojiri2017}, makes very timely questioning about possible LQG extensions of $f(\mathcal{R})$ models, especially via their scalar-tensor reformulation \cite{Bergmann1968,Capone2010,Ruf2018}. A first attempt in this direction was performed in \cite{Zhang2011a,Zhang2011,Han2014} (see also \cite{Cianfrani2009b,Bombacigno2018,Bombacigno2019}), where the problem was faced by considering the metric as the only independent field, and authors actually followed in defining Ashtekar-like variables an extended phase-space method \cite{Thiemann2007}. Conclusions of this study suggest that a suitable set of variables can be determined, and Gauss and vector constraints properly obtained, with the non-minimally coupled scalar field affecting the scalar constraint.
\\ \indent Here, we face the same problem on a more general framework and adopting the most natural first order formalism, i.e. we deal with Palatini $f(\mathcal{R})$ models \cite{Olmo2011}, equipped with Holst and Nieh-Yan terms. In particular, we characterize the resulting classical theory and we discuss the reformulation in terms of $SU(2)$ variables. 
\\ \indent In including Holst and Nieh-Yan terms, we have two different choices, consisting in inserting these terms either inside or outside the argument of the function $f$. We first analyze the classical dynamics of these models demonstrating that they correspond to two physically distinct scenarios. In fact, both when the Holst term is included in the argument of $f$ and when the Nieh-Yan term is added to the $f(\mathcal{R})$ Lagrangian, we recover the dynamics of Palatini $f(\mathcal{R})$ models, characterized by a non-dynamical scalar field. Conversely, when the Holst term is added to $f$ and the Nieh-Yan one is plugged inside the function, we deal with a scalar-tensor theory, where the kinetic term for the scalar field is modulated by the Immirzi parameter. We specialize, then, to this case, showing how a non vanishing value of the Immirzi parameter causes the scalar field to acquire an independent dynamical character. We perform the Hamiltonian formulation and we discuss the resulting morphology in terms of the constraints emerging after the Legendre transformation.\\
The main merit of this study consists of the determination of suitable generalized Ashtekar-Barbero-Immirzi variables starting from a genuine first order action, and we determine Gauss and vector constraints with the same form of LQG, i.e. we are able to construct a kinematic Hilbert space suitable for LQG canonical quantization. We study the spectrum of the area operator, and differently from what assumed in \cite{Fatibene2010}, we clarify how the area operator has an unambiguous geometrical nature, being constructed with the real triad of the space. We stress, therefore, how the different link between the real triads and the particular $SU(2)$ variables considered affects the morphology of the area operator. In the case of a standard Palatini $f(\mathcal{R})$ model, the field is non-dynamical and when it is expressed via the trace of the stress energy tensor, a coupling takes place among the size of the area associated to a graph and the nature of the matter filling the space. More interesting is when the scalar field is truly dynamical, and it must be quantized as a proper scalar degree of freedom. In this case, following \cite{Lewandowski2016}, the discrete character of the area is spoiled by the continuous spectrum of the scalar field.\\
In this regard, we can infer that in extended theories we considered the ambiguity of Immirzi parameter is to some extent weakened. Our study, indeed, suggests that such a parameter could be reabsorbed in the scalar field definition, hinting a more general formulations of the gravitational sector as a $SU(2)$ gauge theory. We emphasize that the quantization procedure for the scalar field requires in both the cases a very particular attention. Indeed, when non dynamical, it still relies on the quantization of the truly gravitational degrees of freedom, which are of course involved in the very definition of the trace for the stress energy tensor. On the other hand, the dynamical case is sensitive to the non minimal coupling of the scalar field and the $\phi$-factors appearing in the Hamiltonian constraint must be treated carefully.\\
Furthermore, when comparing the second order metric analysis of \cite{Zhang2011} to our Palatini formulation, we observe the emergence of a discrepancy in the scalar constraints. By other words, starting directly from a metric formulation of $f(R)$ gravity with $SU(2)$ variables provides different dynamical constraints with respect to a first order formulation. In this respect, we outline the possibility to restore a complete equivalence between these two approaches, by restating our models into the Einstein framework and performing a canonical transformation. We note that similar issues hold also for \cite{Zhou2013}, in which the analysis is actually pursued in a first order formalism, starting from an action which differs however from the one considered here by the inclusion of additional contributions which eliminate torsion from the theory.
\\ \indent The paper is structured as follows. In section~\ref{sec2} the models are presented and their effective theories are derived, highlighting equivalences and differences with Palatini $f(\mathcal{R})$ theory in its scalar-tensor formulation. In section~\ref{Appendix} the spacetime splitting and Hamiltonian analysis of the constrained system are performed. In section~\ref{Sec mod asht var} the new modified variables and the correspondent set of constraints are derived. In particular, we depict some possible effects on the spectrum of the area operator in the presence of a scalar field, and we also briefly discuss the case when it is devoid of a proper dynamics. The analysis performed in the conformal Einstein frame and the comparison with earlier studies in literature are contained in section~\ref{Sec Einstein frame}. In section~\ref{sec concl} conclusions are drawn, while some details regarding the results presented in section~\ref{Appendix} can be found in the~\ref{new Appendix}.
\\ \indent Eventually, notation is established as follows. Spacetime indices are denoted by middle alphabet Greek letters $\mu,\nu,\rho$, spatial ones by letters from the beginning of the Latin alphabet $a,b,c$. Four dimensional internal indices are displayed by capital letters from the middle of the Latin alphabet $I,J,K$, while $i,j,k$ indicate three dimensional internal indices. Spacetime signature is chosen mostly plus, i.e. $\eta_{\mu\nu}= \text{diag}(-1,1,1,1)$.
\section{Extended $f(\mathcal{R})$ actions}\label{sec2}
\noindent We consider the following models
\begin{align}
S &= \frac{1}{2\chi} \int d^4x \sqrt{-g} f(\mathcal{R}+L), \label{fRH} \\
S&= \frac{1}{2\chi} \int d^4x \sqrt{-g} \left[ f(\mathcal{R})+L \right],\label{fR H}
\end{align}
where $\chi=8\pi G$.  The Ricci scalar $\mathcal{R}=g^{\mu\nu}\mathcal{R}_{\mu\nu}$ is obtained by the contraction of the Ricci tensor $\mathcal{R}_{\mu\nu}$, here considered as a function of the independent connection $\Gamma\indices{^{\mu}_{\nu\rho}}$ and related to the Riemann tensor by $\mathcal{R}_{\mu\nu} = \mathcal{R}\indices{^{\rho}_{\mu\rho\nu}}$, with $\mathcal{R}\indices{^{\mu}_{\nu\rho\sigma}} = \partial_{\rho}\tensor{\Gamma}{^{\mu}_{\nu\sigma}} - \partial_{\sigma}\tensor{\Gamma}{^{\mu}_{\nu\rho}} + \tensor{\Gamma}{^{\mu}_{\lambda\rho}}\tensor{\Gamma}{^{\lambda}_{\nu\sigma}} - \tensor{\Gamma}{^{\mu}_{\lambda\sigma}}\tensor{\Gamma}{^{\lambda}_{\nu\rho}}$. The term $L$ either coincides with the Holst term or with the Nieh-Yan invariant, given by, respectively
\begin{align}
    H & = -\frac{\beta}{2}\varepsilon\indices{^{\mu\nu\rho\sigma}}\mathcal{R}_{\mu\nu\rho\sigma},
    \label{holst def}\\
    NY & = \frac{\beta}{2}\varepsilon\indices{^{\mu\nu\rho\sigma}}\left( \frac{1}{2} T\indices{^{\lambda}_{\mu\nu}}T\indices{_{\lambda\rho\sigma}} - \mathcal{R}_{\mu\nu\rho\sigma} \right),
    \label{ny def}
\end{align}
with $\beta$ the reciprocal of the Immirzi parameter. Torsion tensor is displayed by $T\indices{^{\mu}_{\nu\rho}}=\Gamma\indices{^{\mu}_{\nu\rho}} - \Gamma\indices{^{\mu}_{\rho\nu}} $ and, as outlined in \cite{Bombacigno2018,Bombacigno2019,Calcagni2009}, it cannot be a priori neglected in Palatini $f(\mathcal{R})$ generalizations of Holst and Nieh-Yan actions, being its form to be determined dynamically.\\
It is worth noting that by dealing with a proper metric-affine formalism, the affine connection $\Gamma\indices{^\mu_{\nu\rho}}$ can be a priori characterized by non-metricity as well, measuring the departure from metric compatibility, i.e. $Q\indices{_{\rho\mu\nu}}=-\nabla_\rho g_{\mu\nu}\neq 0$. However, it can be demonstrated (see \cite{Iosifidis:2019fsh,Iosifidis:2020dck,Jimenez:2020dpn} for details) that by means of the projective transformation
\begin{equation}
    \Gamma\indices{^\rho_{\mu\nu}}\to \Gamma\indices{^\rho_{\mu\nu}}+\delta\indices{^\rho_\mu}\xi_\nu,
\end{equation}
where $\xi_\nu$ is a vector degree, non-metricity contributions can be always neglected, so that we can safely assume connection be still metric compatible. Therefore, without loss of generality, we can rewrite the connection as $\tensor{\Gamma}{^{\rho}_{\mu\nu}} = \tensor{C}{^{\rho}_{\mu\nu}} + \tensor{K}{^{\rho}_{\mu\nu}}$,
where $\tensor{C}{^{\nu}_{\rho\mu}}$ denotes the Christoffel symbol of $g_{\mu\nu}$ and the independent character of the connection is now encoded in the contortion tensor, given by $\tensor{K}{^{\rho}_{\mu\nu}}=\frac{1}{2}(\tensor{T}{^{\rho}_{\mu\nu}} - \tensor{T}{_{\mu}^{\rho}_{\nu}} - \tensor{T}{_{\nu}^{\rho}_{\mu}})$. Now, in complete analogy with $f(R)$ theories, the actions \eqref{fRH} and \eqref{fR H} can be expressed in the Jordan frame as
\begin{align}\label{Jordan frame models}
    S &= \frac{1}{2\chi} \int d^4x \sqrt{-g} \left[  \phi \left(\mathcal{R}+L\right) - V(\phi) \right],\\
    S &= \frac{1}{2\chi} \int d^4x \sqrt{-g} \left[  \phi \mathcal{R} +L - V(\phi) \right],\label{Jordan frame models 2}
\end{align}
with $\phi$ defined as the derivative of the function $f(\cdot)$ with respect to its generic argument, while the potential takes the same expression as in standard Palatini $f(\mathcal{R})$ theory.\\
Next, the effective theories dynamically equivalent on-half shell to models \eqref{Jordan frame models} and \eqref{Jordan frame models 2}, can be computed inserting into the actions the solution of the equations of motion for the independent connection. This can be easily achieved decomposing the torsion tensor into its independent components according to the Lorentz group. These are the trace vector $T_{\mu} \equiv T \indices{^{\nu}_{\mu\nu}}$, the pseudotrace axial vector
$S_{\mu} \equiv \varepsilon_{\mu\nu\rho\sigma}T^{\nu\rho\sigma}$
and the antisymmetric tensor $q_{\mu\nu\rho}$, satisfying $
\varepsilon^{\mu\nu\rho\sigma} q_{\nu\rho\sigma} = 0$ and $ q\indices{^{\mu}_{\nu\mu}} = 0$. Their equations of motion can be solved yielding $q_{\mu\nu\rho}\equiv 0$, whereas vectors $S_{\mu}$ and $T_{\mu}$ can be expressed in terms of $\partial_{\mu}\phi$ as
\begin{align}\label{tor sol t}
&T_\mu = \frac{3}{2\phi}\left[ \dfrac{1 + b_1b_2\Phi\beta^2/\phi}{1 + b_2\Phi^2\beta^2 /\phi^2} \right]\partial_{\mu}\phi,\\
&S_\mu=\frac{6\beta}{\phi} \left[ \dfrac{b_1-b_2\Phi/\phi}{1 + b_2\Phi^2\beta^2/\phi^2} \right]\partial_{\mu}\phi.\label{tor sol s},
\end{align}
where we introduced two parameters $b_1$ and $b_2$ which can take values $0$ or $1$ according to what specific model considered. If the Holst term is taken into account, then $b_2=1$, while $b_1=0$ and $b_1=1$ for $f(\mathcal{R})+H$ and $f(\mathcal{R}+H)$, respectively. When the action features the Nieh-Yan contribution, $b_2=0$, while $b_1=0$ and $b_1=1$ for $f(\mathcal{R})+NY$ and $f(\mathcal{R}+NY)$. Finally, $\Phi$ is coincident with $\phi$ in the $f(\mathcal{R}+H)$ case and identically equal to $1$ in the $f(\mathcal{R})+H$ one.\\
Substituting these results back into the actions yields the effective theory
\begin{equation}
S=\frac{1}{2\chi} \int  d^4x \sqrt{-g} \left[ \phi R - \frac{\Omega(\phi)}{\phi}\partial^{\mu}\phi\partial_{\mu}\phi - V(\phi) \right],
\label{effective brans dicke}
\end{equation}
where $\Omega(\phi)$ depends on the particular model addressed. Both for $f(\mathcal{R}+H)$ and $f(\mathcal{R})+NY$ it assumes the constant value $\Omega=-3/2$, corresponding to the effective description of standard Palatini $f(\mathcal{R})$ gravity. This implies that the field $\phi$ is not an actual degree of freedom, but it is simply determined by the structural equation as in the standard case \cite{Sotiriou2010,Olmo2011}, i.e.
\begin{equation}
2V(\phi)-\phi V'(\phi)=\chi T,
\label{structural equation}
\end{equation}
where a prime denotes differentiation with respect to the argument and $T$ is trace of the stress energy tensor for some matter contributions, which are assumed do not couple to connection. Thus, as one might expect, the topological character of the Nieh-Yan term is preserved if it is added directly to the Lagrangian. Less trivial, instead, is of course the outcome for the $f(\mathcal{R}+H)$ model. In this case, indeed, the vanishing of the Holst term on half shell is to some extent recovered when it is included in the argument of the function $f(\cdot)$.\\
On the other hand, inserting the Nieh-Yan term in the argument of the function $f(\cdot)$ or featuring the Palatini $f(\mathcal{R})$ theory with an additional Holst contribution, leads to the following expressions, respectively:
\begin{equation}
\Omega=\frac{3(\beta^2-1)}{2},\quad
\Omega=-\frac{3}{2}\frac{\phi^2}{\phi^2 + \beta^2},
\end{equation}
ensuring the classical equivalence to Palatini $f(\mathcal{R})$ gravity for $\beta=0$. In this case, therefore, the scalar field acquires in general a dynamical character due to a non vanishing value of the Immirzi parameter, and \eqref{structural equation} is replaced by
\begin{equation}
    (3+2\Omega)\Box\phi+\Omega'(\partial\phi)^2+2V(\phi)-\phi V'(\phi)=\chi T.
    \label{scalar equation phi}
\end{equation}
Now, by virtue of the dependence of $T_\mu,\,S_\mu$ on derivatives of scalar field $\phi$, we actually deal with a theory equipped with propagating torsion degrees, by close analogy with \cite{Bombacigno2018,Bombacigno2019}, where $f(\mathcal{R})+H$ and $f(\mathcal{R})+NY$ models were considered in the presence of a dynamical Immirzi field. As a result, even though the $f(\mathcal{R}+NY)$ model is formally identical to \textit{metric} $f(R)$ gravity for $\beta=\pm 1$  ($\Omega = 0$), they are actually endowed with distinct phenomenology.

\section{Analysis of the constrained Hamiltonian system}\label{Appendix}
In this section we perform the spacetime splitting and Hamiltonian analysis of the models presented above, with the aim of characterizing the phase space structure of the theory.\\
Since we are eventually interested in the implementation of Ashtekar-like variables, let us first recall the two possible procedures available to achieve this result, namely the so called extended phase space approach, which was followed in [30,31,32], and the first order approach based on the addition of Holst or Nieh-Yan contributions, which is the one we adopted in this paper, consistently with the Palatini formulation of $f(R)$ gravity.\\
The former, the so called extended phase space approach, is carried out in a second order formalism at the Hamiltonian level, firstly defining new configuration variables, i.e. the extrinsic curvature and densitized triad as in standard LQG, and then by extending the phase space to a larger one, characterized by an additional constraint. Next, a symplectic reduction is performed, showing that the new phase space reproduces the correct Poisson brackets between the old configuration variables, namely the 3-metric and its conjugate momentum. Finally, the Ashtekar variables are introduced by means of a canonical transformation on the new phase space variables. This is the paradigm followed in \cite{Zhang2011a,Zhang2011,Han2014}. \\
The other approach is pursued directly at Lagrangian level according a first order or Palatini formalism, by including additional terms in the action, such as the Holst or Nieh-Yan contributions. The formulation in terms of densitized triads and extrinsic curvature arises  naturally after the 3+1 spacetime decomposition and Legendre transform are performed. Eventually, with the same canonical transformation, a constrained Hamiltonian system coordinatized by Ashtekar variables is obtained.\\
In this paper we follow this very last approach for both models \eqref{fRH}-\eqref{fR H}.\\
Let us start performing the spacetime splitting on actions \eqref{Jordan frame models} and \eqref{Jordan frame models 2}, which, by means of tetrad fields $e^I_{\mu}$ and spin connections $\omega\indices{_{\mu}^{IJ}}$, can be simultaneously rewritten, modulo surface terms, as
\begin{align}\label{Appendix eq A1}
&S= \dfrac{1}{2\chi} \int d^4x \; e \left[\phi e^{\mu}_I e^{\nu}_J R^{IJ}_{\mu\nu} + \dfrac{\phi}{24}S^{\mu}S_{\mu} -\dfrac{2}{3}\phi T^{\mu}T_{\mu} \right. \\
&
\left. +2T^{\mu}\partial_{\mu}\phi - b_1 \dfrac{\beta}{2}S^{\mu}\partial_{\mu}\phi  + b_2\Phi \frac{\beta}{3}T^{\mu}S_{\mu} - V(\phi) \right],\nonumber
\end{align}
where $e=\text{det}(e^I_{\mu})$ and $R^{IJ}_{\mu\nu} = 2\partial_{[\mu} \omega\indices{_{\nu]}^{IJ}} + 2\omega\indices{_{[\mu}^I_{K}}\omega\indices{_{\nu]}^{KJ}}$ is the strength tensor of the spin connection. In \eqref{Appendix eq A1}, terms containing $q_{\mu\nu\rho}$ have been neglected since they would eventually turn out to yield vanishing contributions as argued further on.\\
The spacetime splitting is achieved via a foliation of the manifold into a family of 3-dimensional hypersurfaces $\Sigma_t$ defined by the parametric equations $y^{\mu} = y^{\mu}(t,x^a)$, where $t\in \mathbb{R}$. The submanifold $\Sigma_t$ is globally defined by a time-like vector $n^{\mu}$ normal to the hypersurfaces, such that $n^{\mu}n_{\mu}=-1$, and an adapted base on $\Sigma_t$ is then given by $b^{\mu}_a := \partial_{a}y^{\mu}$, satisfying the conditions $g_{\mu\nu}n^{\mu}b^{\nu}_a=0$. Defining the deformation vector as $t^{\mu}=\partial_{t}y^{\mu}$, it can be decomposed on the basis vectors $\left\lbrace n^{\mu},b^{\mu}_a \right\rbrace$ as $t^{\mu} = N n^{\mu} + N^{\mu}$, where $N^{\mu}=N^{a} b^{\mu}_a$, $N$ is the lapse function and $N^a$ the shift vector.\\
The completeness relation $h_{\mu\nu} = g_{\mu\nu} + n_{\mu}n_{\nu}$ holds, where $h_{\mu\nu} = h_{ab} b^a_{\mu} b^{b}_{\nu}$ is the projector on the spatial hypersurfaces and $h_{ab}$ the 3-metric, related to the triads by $h_{ab} = e^i_a e_{ib}$. The lapse function, shift vector and the 3-metric are the new metric configuration variables, in terms of which the metric acquires the usual ADM expression, i.e.
\begin{equation}
ds^2=-N^2 dt^2 + h_{ab}(dx^a+N^adt)(dx^b+N^b dt),
\end{equation}
as in standard geometrodynamics.\\
Now, assuming the time gauge conditions $n^{\mu}=e^{\mu}_0$, $e^t_i = 0$ and using $e=N\bar{e}$, with  $\bar{e}=\text{det}(e^i_a)$, the action can be rewritten as
\begin{align}\label{action Sb}
    S_{b} &=  \int dt d^3x \; \bar{e} \left\lbrace \phi e^a_i \left[  \mathcal{L}_t K^i_a - D^{(\omega)}_a (t\cdot \omega^i) + t\cdot \tensor{\omega}{^{i}_{k}}K^k_a\right] + \right. \nonumber \\&
   + \left( - n\cdot T +b_1\frac{\beta}{4}n\cdot S \right) \mathcal{L}_t \phi
   -N^a \left[ 2 \phi e^b_i D^{(\omega)}_{[a}K^i_{b]} +\left(- n\cdot T + b_1\frac{\beta}{4}n\cdot S \right)\partial_a\phi \right]+
   \nonumber \\ &
   +N\left[ \frac{\phi e^a_i e^b_j}{2} \left( {}^3R^{ij}_{ab} + 2K^i_{[a}K^j_{b]} \right) + \left(T^a - b_1\frac{\beta}{4}S^a\right)\partial_{a}\phi  -\frac{\phi}{48}(n\cdot S)^2 + \frac{\phi}{3}(n\cdot T)^2 + \right.\nonumber\\& \left.\left.
   -b_2\Phi\frac{\beta}{6}(n\cdot T)(n \cdot S)  +\frac{\phi}{48}S^aS_a - \frac{\phi}{3}T^aT_a + b_2 \Phi \frac{\beta}{6}T^aS_a  -\frac{1}{2}V(\phi) \right]  \right\rbrace,
\end{align}
where $K^i_a\equiv\omega^{0i}_a$, the Lie derivative along the vector field $t^{\mu}$ is defined as $\mathcal{L}_t V_{\mu} = t^{\nu}\partial_{\nu}V_{\mu} + V_{\nu}\partial_{\mu}t^{\nu}$, while $\cdot$ indicates spacetime indices contractions, namely $t\cdot \omega^i\equiv t^{\mu}\omega^{0i}_{\mu}$, $t \cdot \omega^{ij}\equiv t^{\mu}\omega^{ij}_{\mu}$, $n \cdot T \equiv n^{\mu}T_{\mu}$ and $n \cdot S \equiv n^{\mu}S_{\mu}$. Moreover, we defined the derivative $D^{(\omega)}_a$ acting only on spatial internal indices via the spatial components of the spin connection, i.e. $D^{(\omega)}_a V^i_{b}= \partial_a V^i_b + \omega\indices{_a^i_j}V^j_b$. Then, the computation of conjugate momenta of $K^i_a$ and $\phi$ yields, respectively
\begin{align}
\tilde{E}^a_i &\equiv \dfrac{\delta S_{tot}}{\delta \mathcal{L}_t K^i_a} = \phi \bar{e} e^a_i,\\
\pi &\equiv \dfrac{\delta S_{tot}}{\delta \mathcal{L}_t \phi} = \bar{e} \left(-  n \cdot T + b_1\frac{\beta}{4} n \cdot S \right),
\end{align}
while all other momenta vanish. Thus, all momenta are non invertible for the correspondent velocities, implying the presence of just as many primary constraints (See the \ref{new Appendix} for the detailed list of primary constraints). \\
A total Hamiltonian can be therefore defined by replacing each non invertible velocity with a Lagrange multiplier, i.e.:
\begin{equation}\label{Htot}
\begin{aligned}
&H_T =\int d^3x \left( \tilde{E}^a_i \mathcal{L}_t K^i_a + \pi \mathcal{L}_t \phi +  \lambda^m C_m - L \right)=\\
&=  \int d^3x \left\lbrace 
-t\cdot \omega^i D^{(\omega)}_a (\tilde{E}^a_i) + 
 t\cdot \tensor{\omega}{^{i}^{k}}K_{ai}\tilde{E}^a_k + \right.\\
 &\left. N^a \left[ 2 \tilde{E}^b_i D^{(\omega)}_{[a}K^i_{b]} + \sqrt{\frac{\tilde{E}}{\phi^3}} \left( -n\cdot T + b_1\frac{\beta}{4} n \cdot S \right) \partial_a\phi \right]  \right.\\&
 \left.-N \sqrt{\phi} \frac{\tilde{E}^a_i\tilde{E}^b_j}{2\sqrt{\tilde{E}}} \left( {}^3R^{ij}_{ab} + 2K^i_{[a}K^j_{b]} \right) \right. \\&
\left. - N\sqrt{\frac{\tilde{E}}{\phi^3}}\left[ \left(T^a - b_1\frac{\beta}{4}S^a\right)\partial_{a}\phi  - \frac{\phi}{48}(n\cdot S)^2  \right.\right.+ \\&
\left.\left.
 + \frac{\phi}{3}(n\cdot T)^2- b_2\frac{\Phi\beta}{6}(n\cdot T)(n \cdot S) +\frac{\phi}{48}S^aS_a \right.\right.\\
 &\left.\left. - \frac{\phi}{3}T^aT_a + b_2 \Phi\frac{\beta}{6}T^aS_a  - \frac{1}{2} V(\phi) \right] +\lambda^m C_m \right\rbrace
\end{aligned}
\end{equation}
where $\tilde{E}=\text{det}(\tilde{E^a_i})$ and $\lambda^m C_m$ in the first line collectively indicates the primary constraints and their correspondent Lagrange multipliers, indicated by $\lambda$ characters (see the \ref{new Appendix}). Finally, the phase space is equipped with the standard Poisson brackets, which are defined in the Appendix as well. \\
At this stage we have to impose that primary constraints be preserved by the dynamics of the system. This amounts to compute their time evolution, evaluating their Poisson brackets with the total Hamiltonian using \eqref{Poisson1}-\eqref{Poisson2}, and imposing the result to be at least weakly vanishing on the constraint hypersurface. This yields
\begin{align}\label{kCdot}
{}^{(K)}\dot{C}^a_i &= Z^a_i - \f\bar{e} \Delta^{aj}_{bi}  {}^{(e)}\lambda^b_j  \approx0,\\\label{Cdot}
\dot{C}&=W  + \bar{e} \,{}^{(T)}\lambda - \bar{e}\, b_1\frac{\beta}{4}  {}^{(S)}\lambda\approx0,\\\label{eCdot}
{}^{(e)}\dot{C}^i_a &= \f \bar{e} \Delta^{bi}_{aj} {}^{(K)}\lambda^j_b\approx0,\\\label{omegatCdot}
{}^{(\omega_t)}\dot{C}_i&= D^{(\omega)}_a \E^a_i\approx0,\\\label{omegaaCdot}
{}^{(\omega_a)}\dot{C}^a_{ij}  &= t\cdot \omega^{[i} \E^{j]a} - 2 N^{c}\E^{a}_{[i}K_{j]c} - N^a \E^c_{[i} K_{j]c}  \nonumber\\
& - \frac{\E^b_{[i}\E_{j]a}}{\sqrt{\E}}\partial_b \left( N\sqrt{\phi} \right) - N\sqrt{\phi} D^{(\omega)}_b \left( \frac{\E^b_{[i}\E_{j]a}}{\sqrt{\E}} \right)\approx0,\\\label{NaCdot}
{}^{(N)}\dot{C}_a &= - \left( 2 \E^b_i D^{(\omega)}_{[a}K^i_{b]} + \pi\partial_a\phi \right)
  \equiv H_a \approx0,\\\label{NCdot}
 {}^{(N)}\dot{C} &=   \sqrt{\phi} \frac{\E^a_i\E^b_j}{2\sqrt{\E}} \left( {}^3R^{ij}_{ab} + 2K^i_{[a}K^j_{b]} \right) + \sqrt{\frac{\E}{\phi^3}}\left[ \left(T^a - b_1 \frac{\beta}{4}S^a\right)\partial_{a}\phi  \right. \nonumber \\&
  - \frac{\phi}{48}(n\cdot S)^2 + \frac{\phi}{3}(n\cdot T)^2  - b_2 \Phi\frac{\beta}{6}(n\cdot T)(n \cdot S)  +\frac{\phi}{48}S^aS_a \nonumber\\&
\left.- \frac{\phi}{3}T^aT_a   + b_2 \Phi \frac{\beta}{6}T^aS_a  - \frac{1}{2} V(\phi) \right]    \equiv -H\approx0,\\\label{rotCdot}
{}^{(\omega_t)}\dot{C}_{ik} &= K_{a[k}\E^a_{i]}\approx0,\\\label{SaCdot}
{}^{(S)}\dot{C}^a &= N\sqrt{\frac{\E}{\f^3}} \left( -b_1\frac{\beta}{4}\partial^{a}\f + \frac{\f}{24}S^a + b_2\frac{\Phi\beta}{6}T^a \right)\approx0,\\\label{TaCdot}
{}^{(T)}\dot{C}^a &= N\sqrt{\frac{\E}{\f^3}} \left( \partial^{a}\f -\frac{2}{3}\f T^a + b_2 \frac{\Phi\beta}{6}S^a \right)\approx0,\\\label{SCdot}
{}^{(S)}\dot{C} &= -N\sqrt{\frac{\E}{\phi^3}} \left( \frac{\f}{24}n\cdot S  + b_2\frac{\Phi\beta}{6} n\cdot T   \right) +\bar{e} \lambda b_1 \frac{\beta}{4}\approx0,\\\label{TCdot}
{}^{(T)}\dot{C}&=  N\sqrt{\frac{\E}{\phi^3}} \left( \frac{2\f}{3}n\cdot T - b_2\frac{\Phi\beta}{6}n\cdot S \right) -\bar{e} \lambda\approx0,
\end{align}
where we defined
\begin{align}
Z^a_i &\equiv\left\lbrace \E^a_i , H_T \right\rbrace  
- \bar{e} e^a_i \left\lbrace  \phi  , H_T  \right\rbrace, \\
\Delta^{aj}_{bi} &\equiv \left( \delta^a_b \delta^j_i - e^a_i e^j_b  \right),\\
W &\equiv\left\lbrace \pi , H_T \right\rbrace + \left(   n\cdot T   - b_1\frac{\beta}{4} n\cdot S \right) \left\lbrace \bar{e} , H_T \right\rbrace.
\end{align}
The expressions above are either functions of the phase space variables alone or they also depend on the Lagrange multipliers. In the former case, the weakly vanishing of the expressions must be imposed, resulting in the presence of secondary constraints. In the latter, they have to be considered as equations in the Lagrange multipliers and must be solved for them, restricting their original arbitrariness.\\
Before moving on, we note that some phase space variables, namely $e^a_i$, $t\cdot \omega^i$, $\omega\indices{_a^{ij}}$, $N^a$, $N$, $t\cdot \omega^{ij}$, $S_a$, $T_a$, $n\cdot S$, $n\cdot T$, are actually completely arbitrary and can be considered as Lagrange multipliers themselves. This happens because their momenta only appear in the combination $\lambda^m C_m$ in $H_T$ and therefore their dynamics is itself arbitrary, being given only by the correspondent Lagrange multiplier.\\
Now, expressions \eqref{kCdot}, \eqref{eCdot}, \eqref{omegatCdot}, \eqref{omegaaCdot}, \eqref{SaCdot}, \eqref{TaCdot} contain Lagrange multipliers ${}^{(e)}\lambda^c_k$, ${}^{(K)}\lambda^j_b$, $\omega\indices{_a^{ij}}$, $t\cdot \omega^i$, $S_a$, $T_a$ and are solved by
\begin{align}
{}^{(e)}\lambda^c_k &= \frac{1}{\phi\bar{e}}\left(\Delta^{-1} \right)^{ci}_{ak} Z^a_i,\\
{}^{(K)}\lambda^j_b &= 0,\\\label{solution to omegatCdot}
 \omega\indices{_a^{ij}}&=\tilde{\omega}\indices{_a^{ij}} \equiv\tilde{E}^{b[i} \left( 2 \partial_{[a}\tilde{E}^{j]}_{b]} + \tilde{E}^{j]d} \tilde{E}^k_a \partial_{d}\tilde{E}_{bk}\right) + \frac{1}{\tilde{E}}\tilde{E}^{[i}_a \tilde{E}^{j]b}\partial_{b}\tilde{E},\\
t\cdot \omega^i &= -2N^c K^i_c + \frac{\tilde{E}^{ib}}{\sqrt{\tilde{E}}}\partial_b\left( N\sqrt{\phi} \right),\label{solution to omegaaCdot}
\end{align}
where $(\Delta^{-1})^{ci}_{ak} = \left( \delta^c_a \delta^i_k -\frac{1}{2} e^c_k e^i_a  \right)$ is the inverse of $\Delta^{aj}_{bi}$ defined such that $\Delta^{aj}_{bi}(\Delta^{-1})^{ci}_{ak}= \delta^c_b \delta^j_k$ and the solutions for $S_a$ and $T_a$ are given by the spatial part of expressions \eqref{tor sol t} and \eqref{tor sol s}, which is consistent with the expressions obtained at the Lagrangian level, solving the field equations for torsion.\\
Note that \eqref{solution to omegaaCdot} is a solution to \eqref{omegaaCdot} since in the latter the third term is proportional to the rotational constraint, which will be derived shortly, while the last term can be dropped once the spin connection is set as in \eqref{solution to omegatCdot}.\\
Eventually, the spatial components of the spin connection turn out to be functions of $\tilde{E}^a_i$. However, this does not imply that we have replaced the initial first order Palatini formulation with a second order one. Indeed, part of the original connection is now encoded in the components of torsion. Moreover, the modified Ashtekar connection ${}^{(\beta)}A^a_i$ obtained in section \ref{Sec mod asht var} contains $K^i_a\equiv \omega\indices{_a^{0i}}$, which is a part of the original spin connection that has not been expressed in terms of any other variable and is still completely independent.\\ 
Equations \eqref{SCdot} and \eqref{TCdot} are solved by
\begin{equation}\label{solution Lagrange mult non-dyn}
    \lambda=\frac{2}{3}\phi N n\cdot T
\end{equation}
and $n \cdot S=0$ in models with $\phi$ non-dynamical, or
\begin{equation}\label{relaz S T NY}
    n \cdot S = 4\beta \; n \cdot T 
\end{equation}
and
\begin{equation}\label{relaz S T H}
    n \cdot S = - \frac{4\beta}{\phi} n \cdot T
\end{equation}
in the $f(\mathcal{R}+NY)$ and $f(\mathcal{R})+H$ cases, respectively.\\
We emphasize that for $f(\mathcal{R})+NY$ and $f(\mathcal{R}+H)$ models we can reproduce the structural equation from the secondary constraint \eqref{Cdot}, provided we fix one between ${}^{(S)}\lambda$ and ${}^{(T)}\lambda$. However, it is worth noting that, as shown in \cite{Olmo2011} for the pure Palatini $f(R)$ case, the same result can be obtained without fixing any of the Lagrange multipliers but making use of the equations of motion and taking into account the scalar constraint.\\
Then, one has to impose the conservation of the structural equation and check if a further constraint arises. This yields a linear homogeneous equation in $\lambda$ which in turn must be proportional to $n\cdot T$ as in \eqref{solution Lagrange mult non-dyn}. Therefore, no tertiary constraints arise and the conservation of the structural equation is just a restriction on the Lagrange multiplier $n\cdot T$, which has to vanish. Thus, the scalar field momentum in non dynamical models turns out to be weakly vanishing, in agreement with the non dynamical character of its conjugate variable $\phi$.\\
In the other two models instead, the time evolution of \eqref{constr phi} can be set to zero fixing ${}^{(S)}\lambda$ or ${}^{(T)}\lambda$, which eventually implies that \textit{both} of them are no longer arbitrary since the equations of motion for $n\cdot S$ and $n\cdot T$ are proportional to their Lagrange multipliers and relations \eqref{relaz S T NY} and \eqref{relaz S T H} hold.
Thus, in this case there are no arbitrary degrees of freedom in the definition of $\pi$ that can be used to freeze its dynamics. Moreover, if one of the Lagrange multipliers is used to reproduce the structure equation also in this case, this would imply the non dynamicity of $\phi$, ending up with an inconsistency and forcing to choose another form for the Lagrange multiplier.\\
Finally, expressions \eqref{NCdot} and \eqref{NaCdot}, which do not contain arbitrary Lagrange multipliers anymore, are imposed to be weakly vanishing implying the presence of the vector, scalar and rotational constraints, namely
\begin{align}\label{vincolo Na sec}
&H_a \equiv - \left( 2 \tilde{E}^b_i D^{(\omega)}_{[a}K^i_{b]} + \pi\partial_a\phi \right) \approx 0,\\
\label{vincolo N sec}
& H \equiv -\sqrt{\phi} \frac{\tilde{E}^a_i\tilde{E}^b_j}{2\sqrt{\tilde{E}}} \left( {}^3R^{ij}_{ab} + 2K^i_{[a}K^j_{b]} \right) + \nonumber \\
& - \sqrt{\frac{\tilde{E}}{\phi^3}}\left[ \left(T^a - b_1 \frac{\beta}{4}S^a\right)\partial_{a}\phi - \frac{\phi}{48}(n\cdot S)^2 + \frac{\phi}{3}(n\cdot T)^2 + \right. \nonumber \\&
   - b_2 \Phi\frac{\beta}{6}(n\cdot T)(n \cdot S)  +\frac{\phi}{48}S^aS_a - \frac{\phi}{3}T^aT_a + \nonumber\\&
\left. + b_2 \Phi \frac{\beta}{6}T^aS_a  - \frac{1}{2} V(\phi) \right]   \approx 0,\\
\label{rot sec}
&K_{a[k}\tilde{E}^a_{i]} \approx 0.
\end{align}
Eventually, taking into account the restrictions on Lagrange multipliers obtained in this section, the above constraints take the form shown in equations \eqref{rotational constraint}, \eqref{vecotr constraint} and \eqref{scalar constraint} of section~\ref{Sec mod asht var}.\\
The algebra of the remaining constraints has already been studied in \cite{Zhang2011}. The treatment applies also to the present case since the two sets of constraints are linked by a canonical transformation, as shown in section \ref{Sec Einstein frame}.\\
Matter can be implicitly included into the theory positing its action to depend only on the metric and the matter fields and not on the connection. Assuming that no primary constraints arise in the matter sector, the constraint structure of the theory gets modified by the addition of the terms $\frac{\delta H_{matt}}{\delta N^a}$ and $\frac{\delta H_{matt}}{\delta N}$ to the vector and scalar constraints \eqref{vincolo Na sec} and \eqref{vincolo N sec}, respectively, being $H_{matt}$ the matter Hamiltonian. In terms of it, from the usual definition of the stress energy tensor of matter in terms of the matter Lagrangian, its trace can be expressed as
\begin{equation}\label{Trace}
T = \frac{2 \phi^{5/2}}{N \sqrt{\tilde{E}}} \left( -\frac{N}{2\phi} \dfrac{\delta H_{matt}}{\delta N} + \dfrac{\delta H_{matt}}{\delta \phi} \right),
\end{equation}
a relation useful in order to recover the structural equation in the Einstein frame formulation developed in section~\ref{Sec Einstein frame}.\\
Finally, let us notice that, if the terms proportional to $q_{\mu\nu\rho}$ would not have been neglected, then, given the absence of derivatives of $q_{\mu\nu\rho}$, its conjugate momentum would have been weakly vanishing. The additional terms proportional to it via a Lagrange multiplier appearing in the total Hamiltonian would have produced secondary constraints whose solutions would have implied in turn the vanishing of $q_{\mu\nu\rho}$ components, since it does not couple to any derivative of the scalar field $\phi$, contrary to what happens to the other components of torsion.

\section{Modified Ashtekar variables}\label{Sec mod asht var}
In this section we successfully implement a modified set of Ashtekar-like variables, still suitable for loop quantization and that ensure the presence of a $SU(2)$ Gauss constraint in the phase space of the theory.\\
We now focus on $f(\mathcal{R})+H$ and $f(\mathcal{R}+NY)$ models. As a result of the analysis pursued in the previous section, the gravitational sector of the phase space turns out to be characterized by the set of canonical variables $\{\pi, \phi;\;\tilde{E}^a_i,K^i_a\}$, where $\pi$ denotes the conjugate momentum to $\phi$ and $\{\tilde{E}^a_i,K^i_a\}$ are defined as
\begin{align}
    K^i_a \equiv \omega\indices{_{a}^{0i}},\label{K}\qquad
    \tilde{E}^a_i \equiv \phi E^a_i,
\end{align}
where $E^a_i = \text{det}(e^j_b) e^a_i$ is the ordinary densitized triad and $\omega\indices{_\mu^{IJ}}$ the independent spin connection. This set of variables is not posited and then justified via a symplectic reduction, as it would be done in the extended phase space approach. Instead, it naturally arises from the spacetime splitting and Legendre transform performed in section~\ref{Appendix} on actions \eqref{Jordan frame models}-\eqref{Jordan frame models 2}.
\\ The phase space is subject to a set of first class constraints consisting of the rotational constraint
\begin{equation}\label{rotational constraint}
R_i \equiv \tensor{\varepsilon}{_{ij}^{k}} K^j_a \tilde{E}^a_k \approx 0,
\end{equation}
the vector constraint
\begin{equation}\label{vecotr constraint}
H_a \equiv 2\tilde{E}^b_i D^{(\tilde{\omega})}_{[a}K^i_{b]} + \pi \partial_{a}\phi \approx 0,
\end{equation}
and the scalar constraint
\begin{align}\label{scalar constraint}
H &\equiv -\frac{\sqrt{\phi}}{2}\frac{\tilde{E}^a_i \tilde{E}^b_j}{\sqrt{\tilde{E}}}\left( {}^3R^{ij}_{ab}(\tilde{\omega}) + 2K^i_{[a}K^j_{b]} \right) + \frac{1}{2}\sqrt{\frac{\phi^{3}}{\tilde{E}}}\frac{\phi}{\Omega}\pi^2 + \nonumber \\
& +\frac{1}{2}\sqrt{\frac{\tilde{E}}{\phi^{3}}}\frac{\Omega}{\phi}\partial^a\phi\partial_a\phi +\frac{1}{2} \sqrt{\frac{\tilde{E}}{\phi^3}}V(\phi) \approx 0.
\end{align}
where $\tilde{E}=\text{det}(\tilde{E}^a_i)$, ${}^{3}R\indices{^{ij}_{ab}}(\tilde{\omega}) = 2\partial_{[a}\tilde{\omega}\indices{_{b]}^{ij}} + 2 \tilde{\omega}\indices{_{[a}^i_k} \tilde{\omega}\indices{_{b]}^{kj}}$. In particular, we defined a new type of covariant derivative $D^{(\tilde{\omega})}_a$, acting on internal spatial indices, by means of the modified spin connection
\begin{equation}\label{spin conn}
   \tilde{\omega}\indices{_a^{ij}} =\tilde{E}^{b[i} \left( 2 \partial_{[a}\tilde{E}^{j]}_{b]} + \tilde{E}^{j]d} \tilde{E}^k_a \partial_{d}\tilde{E}_{bk}\right) + \frac{1}{\tilde{E}}\tilde{E}^{[i}_a \tilde{E}^{j]b}\partial_{b}\tilde{E},
\end{equation}
which can be expressed in terms of the Riemannian spin connection $\bar{\omega}\indices{_\mu^{IJ}}=e^{\nu I}\nabla_\mu e^J_{\nu}$ and the scalar field as $\tilde{\omega}\indices{_a^{ij}} =\bar{\omega}\indices{_a^{ij}}(E) + \frac{1}{\phi} E^{[i}_a E^{j]b}\partial_b\phi$. Then, performing the canonical transformation
\begin{align}\label{mod triads}
\tilde{E}^a_i &\rightarrow  {}^{(\beta)}\tilde{E}^a_i = \beta \tilde{E}^a_i,\\ \label{mod conn}
K^i_a &\rightarrow {}^{(\beta)}A^a_i = \frac{1}{\beta} K^i_a + \tilde{\Gamma}^i_a,
\end{align}
where $\tilde{\Gamma}^i_a =- \frac{1}{2}\varepsilon^{ijk}\tilde{\omega}^{jk}_a$, a set of modified Ashtekar variables $\{{}^{(\beta)}\tilde{E}^a_i,\,{}^{(\beta)}A^i_a\}$ can be obtained, in terms of which the rotational constraint can be combined with the compatibility condition $D^{(\tilde{\omega})}_a\tilde{E}^b_i=0$, satisfied by \eqref{spin conn}, yielding the $SU(2)$ Gauss constraint
\begin{equation}
G_i = \partial_a {}^{(\beta)}\tilde{E}^a_i + \varepsilon_{ijk} {}^{(\beta)}A^j_a {}^{(\beta)}\tilde{E}^a_k \approx 0.
\end{equation}
This guarantees that Palatini $f(\mathcal{R})$ models here considered are actually feasible for LQG quantization procedure. Especially, by means of the new variables the vector constraint can be rearranged as in standard LQG, along with the additional term associated to the scalar field, i.e.
\begin{equation}
    H_a = {}^{(\beta)}\tilde{E}^b_i F^i_{ab} + \pi \partial_a\phi,
\end{equation}
where $F^i_{ab}= 2\partial_{[a} {}^{(\beta)}A^i_{b]} + \varepsilon\indices{^i_{jk}} {}^{(\beta)}A^j_a {}^{(\beta)}A^k_b$. 
Conversely, the scalar constraint turns out to be modified with respect to the standard case, namely
\begin{align}
&H = -\frac{\sqrt{\phi}}{2}\frac{{}^{(\beta)}\tilde{E}^a_i {}^{(\beta)}\tilde{E}^b_j}{\sqrt{\beta{}^{(\beta)}\tilde{E}}}\left[ \beta^2\varepsilon^{ijk}F^k_{ab}+(\beta^2+1) \; {}^3R^{ij}_{ab}(\tilde{\omega}) \right] +  \nonumber \\
& +\frac{1}{2}\sqrt{\frac{\phi^{3}}{{}^{(\beta)}\tilde{E}}}\frac{\phi}{\Omega}\pi^2 +\frac{1}{2}\sqrt{\frac{{}^{(\beta)}\tilde{E}}{\phi^{3}}}\frac{\Omega}{\phi}\partial^a\phi\partial_a\phi +\frac{1}{2} \sqrt{\frac{{}^{(\beta)}\tilde{E}}{\phi^3}}V(\phi),
\label{scalar constraint modified var}
\end{align}
where ${}^{(\beta)}\tilde{E} = \text{det}({}^{(\beta)}\tilde{E}^a_i)$,
reflecting the difference in the dynamics which exists at a classical level between General Relativity and Palatini $f(\mathcal{R})$ Gravity.\\
The preservation of Gauss and vector constraints assure that it is straightforward to extend the usual quantization procedure \cite{Cianfrani2014,Rovelli2004,Thiemann2007} to the new variables \eqref{mod triads}-\eqref{mod conn}, while the quantization of the scalar sector still requires some care, by virtue of the different dynamical character of $\phi$ in the different models. When the scalar field is dynamical, it embodies a proper gravitational degree of freedom and following \cite{Lewandowski2016} we can introduce a scalar field Hilbert space  spanned by quantum states $\ket{\varphi}$, defined by $C^k$-functions $\varphi:\Sigma \rightarrow \mathbb{R}$ and endowed with the scalar product $\braket{\varphi|\varphi}=1,\;\braket{\varphi|\varphi'}=0$, whenever $\varphi\neq\varphi'$. The operator associated to the scalar field acts by multiplication as $\hat{\phi}(x)\ket{\varphi} = \varphi(x)\ket{\varphi}$,
allowing well defined operators $\widehat{\partial_a\phi}$ and $\widehat{\sqrt{\phi}}$, as shown in \cite{Lewandowski2016}. Conversely, the operator associated to the conjugate momentum only exists in its exponentiated version. However, as noticed in \cite{Lewandowski2016}, an operator $\int d^3x f(x) \hat{\pi}(x)$ can still be defined as a directional functional derivative acting in the dual space, spanned by linear functionals $\Phi: C^k(\Sigma)\rightarrow \mathbb{C}$. Now, the main difficulty in the present case comes from the $\phi^{-1}$ factors appearing both in the scalar constraint and in the area (see section~\ref{sec area}). Therefore, in order to define an operator associated to the inverse of the scalar field one can restrict to the case $\phi > 0$ (condition required on a classical level in order the theory be consistent) and resort to the following classical identity
\begin{equation}
\phi^{-1}(x)=4\left( \left\lbrace \sqrt{\phi(x)} , \int d^3z \pi(z) \right\rbrace \right)^2,
\end{equation}
eventually defining the operator $\hat{\phi}^{-1}$ via the replacement $\{\cdot,\cdot\}\rightarrow [\cdot,\cdot]/(i\hslash)$, where braces and square brackets denote Poisson brackets and commutators, respectively.
\subsection{Heuristic picture for the area spectrum}\label{sec area}
In standard LQG the classical expression for the area of a surface $S$ is written in terms of densitized triads as 
\begin{equation}\label{classical area}
    A(S)= \int_S ds \sqrt{E^a_iE^b_i n_a n_b},
\end{equation}
where $n_a$ is the normal vector to the surface, and then quantized computing the action of fluxes on spin-network basis states. The area operator turns out to be diagonal in this basis, with the spectrum given by\footnote{We consider here only the simple case in which there are no nodes of the graph belonging to the surface nor edges laying on it.}
\begin{equation}\label{standard spectrum}
a = \frac{8\pi \ell_P^2}{\beta} \sum_p \sqrt{j_p(j_p+1)},
\end{equation}
where $\ell_P=\sqrt{\hslash G}$ is the Planck length and the sum runs over punctures $p$ of the surface $S$ due to edges of the spin-network, colored by spin quantum numbers $j_p$.\\
However, in the theories we analyzed, the phase space variable to be quantized is ${}^{(\beta)}\tilde{E}^a_i$, whereas the physical metric is still associated to the ordinary densitized triad $E^a_i$. Thus, equation \eqref{classical area} still holds and, in view of its quantization, it has to be rewritten in terms of ${}^{(\beta)}\tilde{E}^a_i$, namely
\begin{equation}\label{mod classical area}
A(S) = \frac{1}{\beta}\int_S ds\, \frac{\sqrt{{}^{(\beta)}\tilde{E}^a_i n_a  {}^{(\beta)}\tilde{E}^b_i n_b}}{\phi} .
\end{equation}
Now, the square root can be quantized via a regularization procedure as in the standard case and, as long as $\phi$ is dynamical, one can treat the reciprocal of the scalar field as explained in section~\ref{Sec mod asht var}. Computing the action of both $\hat{\phi}^{-1}$ and the fluxes on a state obtained by the direct product of spin-network and (dual) scalar field states result in a modified area spectrum
\begin{equation}\label{mod spectrum2}
    a_{\phi} = \frac{8\pi \ell_P^2}{\beta} \sum_p  \dfrac{\sqrt{j_p(j_p+1)}}{\varphi(p)},
\end{equation}
where the only non vanishing contributions are those in which the scalar field is computed in the punctures. This implies that the scalar field contribution to the area operator spoils the discrete character of its spectrum, similarly to what argued in \cite{Veraguth2017,Wang2018}.\\
At the same time, this feature implies that the Immirzi parameter ambiguity, present in standard LQG, is here absent, since different values of $\beta$ do not label different values of physical observables but instead they lead to the same, continuous spectrum. Such an outcome seems to suggest that in Palatini $f(\mathcal{R})$ extensions of LQG, the Immirzi parameter can be conveniently set to unity in definitions of geometrical objects as in \eqref{classical area}, and its effects on dynamics absorbed in the value of $\Omega$. We note that in \cite{Veraguth2017,Wang2018} analogous results are achieved in the context of Conformal-LQG, where an additional conformal transformation is included into the symmetries of the theory. In that case, indeed, it is possible to build an area operator invariant under conformal transformation and independent on the Immirzi parameter, which acquires the  role of gauge parameter for the new conformal symmetry. For such a purpose, however, one is forced to consider the conformal rescaled metric as the physical one, in contrast to our assumptions and by analogy with \cite{Fatibene2010}.
\subsection{Non-dynamical models}
For the two models in which the scalar field is non dynamical, namely $f(\mathcal{R}+H)$ and $f(\mathcal{R})+NY$, the main result presented in this section still holds, namely the existence of a generalized set of Ashtekar variables allowing the preservation of the vector and Gauss constraints.\\
There are however some important caveats. As shown in section~\ref{sec2}, in both cases the parameter $\Omega$ assumes the constant value $\Omega=-3/2$, corresponding to the effective description of standard Palatini $f(\mathcal{R})$ gravity which is characterized by the non-dynamical character of the scalar field.\\
This aspect is classically encoded in the structural equation of motion \eqref{structural equation}. However, we proved that it is also reflected in a slightly different phase space structure. Indeed, the phase space of these models is endowed with an additional second class constraint, which can be recast, provided we properly fix the Lagrangian multipliers, in the form \eqref{structural equation}, proving the non dynamical character of the field $\phi$. This result is reinforced by the fact that its conjugate momentum turns out to be weakly vanishing. In the other two models, instead, no additional constraints arise and the scalar field and its conjugate momentum are truly dynamical.\\
Thus, since in these models the scalar field is not an independent degree of freedom, it has not to be directly quantized, but it has to be considered a function of matter fields $\phi(T)$ by means of \eqref{structural equation}. The same goes for the scalar field appearing in \eqref{mod classical area}. This introduces a dependence of the area operator on matter and we expect that it could affect the purely geometric contribution too, resulting in a modification of the  eigenvalue expression \eqref{standard spectrum} by virtue of the dependence of $T$ on the metric tensor. However, given a matter Lagrangian, even for the simplest choices for the function $f$, the expression for $\phi(T)$ can be quite cumbersome, making its treatment unfeasible.
\section{Formulation in the Einstein frame}\label{Sec Einstein frame}
Here we compare the results of the Hamiltonian approach discussed in section~\ref{Sec mod asht var} to analogous studies present in literature \cite{Zhang2011,Zhou2013}, where a different set of canonical variables was obtained, denoted by hatted characters and related to ours by the canonical transformation
\begin{align}\label{canontransf1}
&\hat{E}^a_i = \frac{1}{\phi}\tilde{E}^a_i,
\qquad\hat{K}^i_a = \phi K^i_a,\\
&\label{canontransf2}
\hat{\pi} = \pi - \frac{1}{\phi} \tilde{E}^a_i K^i_a,
\qquad
\hat{\phi} = \phi.
\end{align}
These results are to some extent controversial, since performing on \eqref{rotational constraint}-\eqref{scalar constraint} such a transformation reproduces the same set of constraints of \cite{Zhang2011,Zhou2013} only if we replace by hand $\Omega$ with  $\Omega + 3/2$. Furthermore, in \cite{Zhang2011} starting from a second order analysis of \eqref{effective brans dicke}, a LQG formulation of scalar tensor theories was achieved via a symplectic reduction technique \cite{Thiemann2007}. Then, this seems to point out that the implementation of Ashtekar variables could be affected by the peculiar choice of the formalism adopted, when extensions to General Relativity are taken into account, as it occurs for the Jordan frame formulation of Palatini $f(\mathcal{R})$ models we considered.
In this sense, the fact that in \cite{Zhou2013} were actually derived results equivalent to \cite{Zhang2011} according a first order approach, can be traced back to the choice of including additional contributions in the action, featuring a Holst term, with the aim of eliminating torsion from the theory.
\\ Now, we want to show how the phase space structure obtained in \cite{Zhang2011,Zhou2013} could be reproduced by means of the canonical transformation \eqref{canontransf1}-\eqref{canontransf2}, provided the analysis of section~\ref{Appendix} be pursued in the so called Einstein frame, endowed with the conformally rescaled metric $\tilde{g}_{\mu\nu} = \phi g_{\mu\nu}$. Of course, we note that such an equivalence does not exclude a priori the existence, even at the level of the Jordan frame, of a different, possibly \textit{gauged} canonical transformation (see \cite{Cianfrani2012}), able to tackle this problem.
\\With steps analogous to the ones followed in the section~\ref{sec2}, the effective actions for the different models can be derived as 
\begin{equation}\label{ST Einstein frame}
S = \frac{1}{2\chi} \int d^4x \sqrt{-\tilde{g}} \left\lbrace \widetilde{R} - \frac{\tilde{\Omega}(\phi)}{\phi}\tilde{g}^{\mu\nu}\partial_{\mu}\phi \partial_{\nu}\phi -U(\phi)\right\rbrace,
\end{equation}
where $\tilde{R}$ is the Ricci scalar depending only on the conformal metric and $U(\phi)=V(\phi)/\phi^2$. As before the models can be divided in two cases. $f(\mathcal{R})+H$ and $f(\mathcal{R})+NY$, for which $\tilde{\Omega}=0$, and $f(\mathcal{R}+NY)$ and $f(\mathcal{R}+H)$, for which $\tilde{\Omega}=\frac{3}{2}\phi \beta^2$ and $\tilde{\Omega}=\frac{3}{2}\frac{\phi\beta^2}{1+\beta^2\phi^2}$, respectively.\\
In particular, by comparing \eqref{ST Einstein frame} with \eqref{effective brans dicke}, we see that $\Omega$ and $\tilde{\Omega}$ are related by
\begin{equation}\label{Jordan Einstein relation}
\tilde{\Omega} = \frac{\Omega + 3/2}{\phi},
\end{equation}
which shows how the value $\Omega=-3/2$, associated to non dynamical configurations for the field $\phi$ in the Jordan frame, corresponds to $\tilde{\Omega}=0$, in agreement with the representation in the Einstein frame of scalar tensor theories.\\
Then, following the line of section~\ref{Appendix}, we can perform the Hamiltonian analysis, which reveals a phase space coordinatized by the set of variables $\{\tilde{E}^a_i,K^i_a\}$, where the densitized triad is now defined in terms of the spatial part of the conformal tetrad $\tilde{e}^I_{\mu} = \sqrt{\phi}e^I_{\mu}$, and a rotational constraint coincident for all models with the expression derived in the Jordan frame. Regarding the vector and scalar constraints instead, in the non dynamical models they read, respectively
\begin{align}
&H_a = 2\tilde{E}^b_i D^{(\tilde{\omega})}_{[a}K^i_{b]}\approx 0\\
&H = -\frac{\tilde{E}^a_i \tilde{E}^b_j}{2\sqrt{\tilde{E}}}\left( {}^3R^{ij}_{ab}(\tilde{\omega}) + 2K^i_{[a}K^
{j}_{b]} \right) + \sqrt{\tilde{E}}U(\phi) \approx 0,
\end{align}
where $\tilde{\omega}\indices{_a^{ij}}$ is the spin connection compatible with $\tilde{e}^i_a$. In this case, the conjugated momentum to the scalar field is weakly vanishing, leading to the primary constraint $\pi\approx 0$, whose conservation along the dynamics simply amounts to impose the variation of the potential term with respect to $\phi$ to weakly vanish. If matter is included this reproduces the structural equation as a secondary constraint, as shown in section \ref{Appendix}.\\
In models where $\phi$ is dynamical, instead, the vector constraint takes the form \eqref{vecotr constraint} and the scalar constraint reads 
\begin{align}
H &= -\frac{\tilde{E}^a_i \tilde{E}^b_j}{2\sqrt{\tilde{E}}}\left( {}^3R^{ij}_{ab}(\tilde{\omega}) + 2K^i_{[a}K^{j}_{b]} \right) + \frac{1}{2\sqrt{\tilde{E}}}\frac{\phi}{\tilde{\Omega}}\pi^2 +\nonumber \\
& + \frac{1}{2}\frac{\tilde{\Omega}}{\phi} \frac{\tilde{E}^a_i\tilde{E}^b_i}{\sqrt{\tilde{E}}} \partial_{a}\phi\partial_b \phi + \sqrt{\tilde{E}}U(\phi) \approx 0.
\end{align}
If we then perform the canonical transformation  \eqref{canontransf1}-\eqref{canontransf2}, the difference consisting in the shift by $-3/2$ of $\Omega$ is now compensated taking into account relation \eqref{Jordan Einstein relation}, and the final expression for the constraints is in agreement with \cite{Zhou2013}, both for $\Omega \neq -3/2$ and for $\Omega = -3/2$ ($\tilde{\Omega}=0$ and $\tilde{\Omega}\neq 0$). Specifically, in the latter case the primary constraint $\pi\approx 0$ becomes $\hat{\pi} + \frac{1}{\hat{\phi}} \hat{E}^a_i \hat{K}^i_a \approx 0$, which reproduces the so called conformal constraint present in \cite{Zhou2013}, and proves the equivalence of first and second order approaches in the Einstein frame.

\section{Concluding remarks}\label{sec concl}
In this work, we adopted the point of view that introducing Ashtekar-Barbero-Immirzi variables into modified $f(\mathcal{R})$ contexts requires a Palatini formulation of the action. This perspective implies the need to add to the Lagrangian the typical Holst and Nieh-Yan contributions, like the standard Einstein-Palatini action in \cite{Holst1996}. As shown in section~\ref{sec2}, we have different possible combinations for the Lagrangian, corresponding to include the aforementioned terms either inside the argument of the function $f$ or simply outside. When the equations of motion are calculated, and the torsion field is expressed via the metric and the scalar field variables, two different physical formulations come out. In fact, plugging the Holst term in the function $f$ or adding the Nieh-Yan contribute directly in the Lagrangian, simply corresponds to a standard Palatini $f(\mathcal{R})$ theory. Conversely, in the opposite case a truly scalar-tensor model is obtained, whose parameter $\Omega$ turns out to depend on the Immirzi parameter.\\
In both instances we are able to define new Ashtekar-Barbero-Immirzi variables, in terms of which Gauss and vector constraints of LQG are recovered. The discrepancy between the two scenarios is also reflected into the morphology of the area operator, obtained starting from its expression in natural geometrical variables, i.e. the natural tetrad fields, and then expressed via the proper $SU(2)$ variables, suitable for loop quantization. As a result, the area spectrum depends now on the scalar field properties as well. In particular, in the case of a Palatini $f(\mathcal{R})$ theory, we have to deal with the intriguing feature that the geometrical structure of the space depends on the nature of the matter by which it is filled. Another interesting issue comes out when the scalar field must be also quantized, and the area operators eigenvalues contain features of the scalar mode spectrum. This property is a consequence of the non-minimal coupling of the scalar field to gravity and suggests space discretization could be influenced by the particular considered form of the function $f$, i.e. of the potential term $V(\phi)$. Thus, the form of the Lagrangian one adopted to describe gravity seems to directly affect the space quantum kinematics, differently from the classical scenario, where only the space metric fixes the geometry kinematics, disregarding the Lagrangian form.\\
We also showed that the scalar field impacts the area spectrum in such a way that it is possible to set the Immirzi parameter equal to one in all the kinematic constraints, while it still affects the scalar constraint morphology.\\
This is not surprising since we are able to directly link the Immirzi parameter to the scalar-tensor parameter $\Omega$. However, the $SU(2)$ morphology of the theory and its kinematic properties must be not influenced by $\Omega$, since we can re-absorbe its value into the Ashtekar-Barbero-Immirzi variable definition. In turn, the dynamics can be instead affected by $\Omega$, i.e. by $\beta$, since the theories are not dynamical equivalent and these parameters can be constrained by experimental observations \cite{Zakharov2006,Chiba2007,Schmidt2008,Berry2011}. \\ In this sense, the Immirzi parameter ambiguity is here completely solved, by its link to the physical scalar-tensor parameter and by the independence of the theory kinematics on its specific value. This is, in our opinion, a very relevant result, opening a new perspective for the solution of some of the LQG shortcomings into a revised and extended formulation of the gravitational interaction.
\\ \indent We conclude by stressing a technical issue concerning the possibility to obtain equivalent formulations, when starting from a second order approach as in \cite{Zhang2011} and according the present Palatini approach. In fact, the scalar constraint appears different in the two analysis and the possibility to restore a complete equivalence implies the choice of a Einstein framework for our formulation, i.e. a conformal rescaling of the tetrad field. This technical evidence suggests a possible physical interpretation for the dynamics in the Einstein framework of a $f(R)$ theory (on the present level it must be considered simply a mathematical tool to make the scalar dynamics minimally coupled), which deserves further investigation.
\section*{Acknowledgements}
The work of F. B. is supported by the Fondazione Angelo della Riccia grant for the year 2020.
\appendix

\section{}\label{new Appendix}
The complete set of conjugate momenta, computed from action \eqref{action Sb}, reads
\begin{align}
K^a_i: \quad &\tilde{E}^a_i \equiv \dfrac{\delta S_{b}}{\delta \mathcal{L}_t K^i_a} = \phi \bar{e} e^a_i,\\
\phi: \quad &\pi \equiv \dfrac{\delta S_{b}}{\delta \mathcal{L}_t \phi} = \bar{e} \left(-  n \cdot T + b_1\frac{\beta}{4} n \cdot S \right);\\
\label{vincolo e}
e^a_i: \quad & {}^{(e)}\pi^i_a \equiv \dfrac{\delta S_{b}}{\delta \mathcal{L}_t e^a_i } = 0; \\
\label{tetrad comp}
t\cdot \omega^i: \quad & {}^{(\omega_t)}\pi_i \equiv \dfrac{\delta S_{b}}{\delta \mathcal{L}_t (t\cdot \omega^i) } = 0;\\
\label{omega}
\omega^{ij}_a: \quad & {}^{(\omega_a)}\pi^a_{ij} \equiv \dfrac{\delta S_{b}}{\delta \mathcal{L}_t (\omega^{ij}_a) } = 0; \\
\label{vincolo Na}
N^a: \quad & {}^{(N)}\pi_a \equiv \dfrac{\delta S_{b}}{\delta \mathcal{L}_t N^a } = 0;\\ 
\label{vincolo N}
N: \quad & {}^{(N)}\pi \equiv \dfrac{\delta S_{b}}{\delta \mathcal{L}_t N } = 0;\\
\label{rot}
t\cdot \tensor{\omega}{^{ik}}: \quad & {}^{(\omega_t)}\pi_{ik} \equiv \dfrac{\delta S_{b}}{\delta \mathcal{L}_t (t\cdot \tensor{\omega}{^{ik}}) } = 0; \\
\label{vincolo Sa}
S_a: \quad & {}^{(S)}\pi^a \equiv \dfrac{\delta S_{b}}{\delta \mathcal{L}_t S_a } = 0; \\
\label{vincolo Ta}
T_a: \quad & {}^{(T)}\pi^a \equiv \dfrac{\delta S_{b}}{\delta \mathcal{L}_t T_a } = 0; \\
\label{vincolo S}
n\cdot S: \quad & {}^{(S)}\pi \equiv \dfrac{\delta S_{b}}{\delta \mathcal{L}_t (n\cdot S) } = 0; \\
\label{vincolo T}
n\cdot T: \quad & {}^{(T)}\pi \equiv \dfrac{\delta S_{b}}{\delta \mathcal{L}_t (n\cdot S) } = 0.
\end{align}
None of the above relation is invertible for the correspondent velocity, yielding the following set of primary constraints
\begin{align}
\label{constr K}
    {}^{(K)}C^a_i &\equiv  \tilde{E}^a_i - \phi \bar{e} e^a_i \approx 0;\\
\label{constr phi}
    C &\equiv  \pi + \bar{e} \left( n\cdot T - b_1\frac{\beta}{4} n \cdot S \right) \approx 0;\\
\label{constr e}
    {}^{(e)}C^i_a &\equiv {}^{(e)}\pi^i_a \approx 0;\\
\label{contsr metr comp}
    {}^{(\omega_t)}C_i &\equiv {}^{(\omega_t)}\pi_i \approx 0;\\
\label{constr conn}
    {}^{(\omega_a)}C^a_{ij} &\equiv  {}^{(\omega_a)}\pi^a_{ij}\approx 0;\\
\label{constr Na}
    {}^{(N)}C_a &\equiv {}^{(N)}\pi_a \approx 0; \\
\label{constr N}
    {}^{(N)}C &\equiv{}^{(N)}\pi \approx 0; \\
\label{constr rot}
    {}^{(\omega_t)}C_{ik} &\equiv {}^{(\omega_t)}\pi_{ik} \approx 0; \\
    \label{constr Sa}
    {}^{(S)}C^a &\equiv {}^{(S)}\pi^a \approx 0;\\
     \label{constr Ta}
    {}^{(T)}C^a &\equiv {}^{(T)}\pi^a \approx 0;\\
     \label{constr S}
    {}^{(S)}C &\equiv {}^{(S)}\pi \approx 0;\\
    \label{constr T}
    {}^{(T)}C &\equiv {}^{(T)}\pi \approx 0.
\end{align}
To enforce each of them in the variational principle one can introduce arbitrary Lagrange multipliers, indicate here by $\lambda$ characters. In particular, they appear in equation \eqref{Htot} via the following expression
\begin{align}
 \lambda^m C_m&= {}^{(K)}\lambda^i_a {}^{(K)}C^a_i +
 \lambda C
 +
{}^{(e)}\lambda^a_i {}^{(e)}C^i_a
+ {}^{(\omega_t)}\lambda^i{}^{(\omega_t)}C_i\\
& 
+ {}^{(\omega_a)}\lambda^{ij}_a{}^{(\omega_a)}C^a_{ij}
+ {}^{(N)}\lambda^a {}^{(N)}C_a 
 + {}^{(N)}\lambda {}^{(N)}C 
 \\&
 + {}^{(\omega_t)}\lambda^{ij}{}^{(\omega_t)}C_{ij} 
  + {}^{(S)}\lambda_a {}^{(S)}C^a 
 +{}^{(T)}\lambda_a {}^{(T)}C^a. \\
 &
  + {}^{(S)}\lambda {}^{(S)}C 
+{}^{(T)}\lambda {}^{(T)}C.
\end{align}
Finally, the phase space is equipped with the following Poisson brackets:
\begin{align}
\label{Poisson1}
\left\lbrace K^i_a (x), \tilde{E}^b_j (y) \right\rbrace &= \delta^i_j \delta^b_a \delta(x,y);\\
\left\lbrace \phi (x), \pi (y)\right\rbrace &= \delta(x,y) ;\\
\left\lbrace e^a_i (x), {}^{(e)}\pi^j_b (y)\right\rbrace &= \delta^j_i \delta^a_b  \delta(x,y);\\
\left\lbrace t\cdot \omega^i (x), {}^{(\omega_t)}\pi_j (y)\right\rbrace &= \delta^i_j \delta(x,y);\\
\left\lbrace \omega\indices{_a^{ij}}(x) , {}^{(\omega_a)}\pi\indices{^b_{kl}} (y)\right\rbrace &= \delta^b_a \delta^{i}_{[k}\delta^{j}_{l]} \delta(x,y);\\
\left\lbrace N^a (x), {}^{(N)}\pi_b(y) \right\rbrace &= \delta^a_b \delta(x,y);\\
\left\lbrace N (x), {}^{(N)}\pi(y) \right\rbrace &= \delta(x,y);\\
\left\lbrace t\cdot\omega^{ij}(x) , {}^{(\omega_t)}\pi_{kl}(y) \right\rbrace &= \delta^{i}_{[k}\delta^{j}_{l]} \delta(x,y,);\\
\left\lbrace S_a (x), {}^{(S)}\pi^b(y) \right\rbrace &= \delta^b_a \delta(x,y);\\
\left\lbrace T_a (x), {}^{(T)}\pi^b(y) \right\rbrace &= \delta^b_a \delta(x,y);\\
\left\lbrace S (x), {}^{(S)}\pi(y) \right\rbrace &= \delta(x,y);\\
\label{Poisson2}
\left\lbrace T (x), {}^{(T)}\pi(y) \right\rbrace &= \delta(x,y).
\end{align}

\bibliographystyle{elsarticle-num}
\bibliography{Bibliography}{}

\end{document}